\documentclass
[twocolumn,prl,unsortedaddress,superscriptaddress,showpacs]{revtex4}%
\usepackage{amsmath}
\usepackage{amsfonts}
\usepackage{amssymb}
\usepackage{graphicx}
\usepackage[utf8]{inputenc}%
\providecommand{\U}[1]{\protect\rule{.1in}{.1in}}
\providecommand{\U}[1]{\protect\rule{.1in}{.1in}}

\begin{document}
\title{Quanta of Geometry: Noncommutative Aspects}
\author{Ali H. Chamseddine}
\affiliation{Physics Department, American University of Beirut, Lebanon}
\affiliation{I.H.E.S. F-91440 Bures-sur-Yvette, France}
\author{Alain Connes}
\affiliation{College de France, 3 rue Ulm, F75005, Paris, France}
\affiliation{I.H.E.S. F-91440 Bures-sur-Yvette, France}
\affiliation{Department of Mathematics, The Ohio State University, Columbus OH 43210 USA}
\author{Viatcheslav Mukhanov}
\affiliation{Theoretical Physics, Ludwig Maxmillians University,Theresienstr. 37, 80333
Munich, Germany }
\affiliation{MPI for Physics, Foehringer Ring, 6, 80850, Munich, Germany}
\email{chams@aub.edu.lb,alain@connes.org,}
\email{viatcheslav.mukhanov@lmu.de}

\begin{abstract}
In the construction of spectral manifolds in noncommutative geometry, a higher
degree Heisenberg commutation relation involving the Dirac operator and the
Feynman slash of real scalar fields naturally appears and implies, by equality
with the index formula, the quantization of the volume. We first show that
this condition implies that the manifold decomposes into disconnected spheres
which will represent quanta of geometry. We then refine the condition by
involving the real structure and two types of geometric quanta, and show that
connected spin-manifolds with large quantized volume are then obtained as
solutions. The two algebras $M_{2}\left(  \mathbb{H}\right)  $ and
$M_{4}\left(  \mathbb{C}\right)  $ are obtained which are the exact
constituents of the Standard Model. Using the two maps from $M_{4}$ to $S^{4}$
the four-manifold is built out of a very large number of the two kinds of
spheres of Planckian volume. We give several physical applications of this
scheme such as quantization of the cosmological constant, mimetic dark matter
and area quantization of black holes.

\end{abstract}

\pacs{0.2.40.Gh,0.2.40.Ky,0.4.60.-m,0.4.70.-s}
\maketitle


\vspace{0.5cm}
\emph{Introduction.---}
To reconcile General Relativity with Quantum Mechanics it is natural to try to
find a generalization of the Hiesenberg commutation relations $[p,q]=i\hbar$.
One expects the role of the momentum $p$ to be played by the Dirac operator,
however, the role of the position variable $q$ is more difficult to discover.
In noncommutative geometry a geometric space is encoded by a spectral triple
$(\mathcal{A},\mathcal{H},D)$ where the algebra $\mathcal{A}$ is the algebra
of functions which interacts with the inverse line element $D$ by acting in
the same Hilbert space $\mathcal{H}$, where $D$ is an unbounded self-adjoint
operator. There is, in the even-dimensional case, an additional decoration
given by the chirality operator $\gamma=\gamma^{\ast}$, $\gamma^{2}=1$,
$D\gamma=-\gamma D$ \cite{AC}. For a compact spin Riemannian manifold $M$ the
algebra $\mathcal{A}$ is the algebra of operator-valued functions on $M$, the
Hilbert space $\mathcal{H}$ is the Hilbert space of $L^{2}$-spinors and the
operator $D$ is the Dirac operator. These operator theoretic data encodes not
only the geometry (the Riemannian metric) but also the $K$-homology
fundamental class of $M$ which is represented by the $K$-homology class of the
spectral triple. Among the operator theoretic properties fulfilled by the
special spectral triples coming from Riemannian geometries, one of them called
the \emph{orientability condition} asserts that, in the even-dimensional case,
one can recover the chirality operator $\gamma$ as an expression of the form
\begin{equation}
\gamma=\sum a_{0}[D,a_{1}]\cdots\lbrack D,a_{n}], \label{orientability}%
\end{equation}
where the $a_{j}\in\mathcal{A}$ and the formal expression is a totally
antisymmetric Hochschild cycle that represents the (oriented) volume form $dv$
of the manifold \cite{Connes} (for simplification, we have dropped the
summation index appearing in this equation). Our goal in this letter is to
show that the quantized Heisenberg commutation relations is a quantized form
of the orientability condition. It was observed in \cite{Connes} that in the
particular case of even spheres the trace of the Chern character of an
idempotent $e$, i.e. $e^{2}=e,$ leads to a remarkably simple operator
theoretic equation which takes the form (up to a normalization factor
$\frac{1}{2^{n/2}n!}$)
\begin{equation}
\left\langle Y\left[  D,Y\right]  \cdots\left[  D,Y\right]  \right\rangle
=\sqrt{\kappa}\,\gamma\quad\left(  n\mathrm{\ terms\,}\left[  D,Y\right]
\right)  \label{yyy}%
\end{equation}
Here $\kappa=\pm1$ and $C_{\kappa}\subset M_{s}(\mathbb{C})$, $s=2^{n/2}$, is
the Clifford algebra on $n+1$ gamma matrices $\Gamma_{A}$, $1\leq a\leq
n+1$\footnote{It is $n+1$ and not $n$ where $\Gamma_{n+1}$ is up to
normalization the product of the $n$ others.}
\[
\Gamma_{A}\in C_{\kappa},\quad\left\{  \Gamma^{A},\Gamma^{B}\right\}
=2\kappa\,\delta^{AB},\ (\Gamma^{A})^{\ast}=\kappa\Gamma^{A}%
\]
We let $Y\in\mathcal{A}\otimes C_{\kappa}$ be of the Feynman slashed form
$Y=Y^{A}\Gamma_{A},$ and fulfill the equations
\begin{equation}
Y^{2}=\kappa,\qquad Y^{\ast}=\kappa Y \label{zzz}%
\end{equation}
When we write $[D,Y]$ in \eqref{yyy}, we mean $[D\otimes1,Y]$. Finally
$\left\langle {}\right\rangle $ applied to a matrix $M_{s}$ of operators is
its trace.

Note that here the components $Y^{A}\in\mathcal{A}$ but it is true in general
that \eqref{zzz} implies that the components $Y^{A}$ are self-adjoint
commuting operators.

In spectral geometry the metric dimension of the underlying space is defined
by the growth of the eigenvalues of the Dirac operator. As shown in
\cite{Connes} for even $n,$ equation \eqref{yyy}, together with the hypothesis
that the eigenvalues of $D$ grow as in dimension $n$, imply that the volume,
expressed as the leading term in the Weyl asymptotic formula for counting
eigenvalues of the operator $D$, is \textquotedblleft
quantized\textquotedblright\ by being equal to the index pairing of the
operator $D$ with the $K$-theory class of $\mathcal{A}$ defined by the
projection $e=(1+\sqrt{\kappa}\,Y)/2$.

In this letter we shall take equation \eqref{yyy}, and its two sided
refinement \eqref{jjj} below using the real structure, as a geometric analogue
of the Heisenberg commutation relations $[p,q]=i\hbar$ where $D$ plays the
role of $p$ (momentum) and $Y$ the role of $q$ (coordinate) and use it as a
starting point of quantization of geometry with quanta corresponding to
irreducible representations of the operator relations. The above integrality
result on the volume is a hint of quantization of geometry. We first use the
one-sided \eqref{yyy} as the equations of motion of some field theory on $M,$
obtained from the spectral data, and describe the solutions as follows. (For
details and proofs see \cite{CCM1,CCM2}).

\emph{Let }$M$\emph{ be a spin Riemannian manifold of even dimension }%
$n$\emph{ and }$(\mathcal{A},\mathcal{H},D)$\emph{ the associated spectral
triple. Then a solution of the one-sided equation exists if and only if }%
$M$\emph{ breaks as the disjoint sum of spheres of unit volume. On each of
these irreducible components the unit volume condition is the only constraint
on the Riemannian metric which is otherwise arbitrary for each component
\cite{CCM1}. }

Each geometric quantum is a topological sphere of arbitrary shape and unit
volume (in Planck units). It would seem at this point that only disconnected
geometries fit in this framework but in the NCG formalism it is possible to
refine \eqref{yyy}. It is the real structure $J$, an antilinear isometry in
the Hilbert space $\mathcal{H}$ which is the algebraic counterpart of charge
conjugation. This leads to refine the quantization condition by taking $J$
into account in the two-sided equation\footnote{The $\gamma$ involved here
commutes with the Clifford algebras and does not take into account an eventual
$\mathbb{Z}/2$-grading $\gamma_{F}$ of these algebras, yielding the full
grading $\gamma\otimes\gamma_{F}$.}
\begin{equation}
\left\langle Z\left[  D,Z\right]  \cdots\left[  D,Z\right]  \right\rangle
=\gamma\quad Z=2EJEJ^{-1}-1,\ \label{jjj}%
\end{equation}
where $E$ is the spectral projection for $\{1,i\}\subset\mathbb{C}$ of the
double slash $Y=Y_{+}\oplus Y_{-}\in C^{\infty}(M,C_{+}\oplus C_{-})$. It is
the classification of finite geometries of \cite{CC2} which suggested to use
the direct sum $C_{+}\oplus C_{-}$ of two Clifford algebras and the algebra
$C^{\infty}(M,C_{+}\oplus C_{-})$. It turns out moreover that in dimension
$n=4$ one has $C_{+}=M_{2}(\mathbb{H}),$ the $2\times2$ matrix algebra whose
elements are quaternions and $C_{-}=M_{4}(\mathbb{C})$ which is in perfect
agreement with the algebraic constituents of the Standard Model \cite{CC2}.
One now gets two maps $Y_{\pm}:M\rightarrow S^{n}$ while, for $n=2,4$,
\eqref{jjj} becomes,
\begin{equation}
\det\left(  e_{\mu}^{a}\right)  =\Omega_{+}+\Omega_{-}. \label{qqq}%
\end{equation}
where $e_{\mu}^{a}$ is the vierbein, with $\Omega_{\pm}$ the Jacobian of
$Y_{\pm}$ (the pullback of the volume form of the sphere).

\emph{ Let }$n=2$\emph{ or }$n=4$\emph{ then \cite{CCM1}, }

$(i)$\emph{~In any operator representation of the two sided equation
\eqref{jjj} in which the spectrum of }$D$\emph{ grows as in dimension }%
$n$\emph{ the volume (the leading term of the Weyl asymptotic formula) is
quantized.}

$(ii)$\emph{~Let }$M$\emph{ be a compact oriented spin Riemannian manifold of
dimension }$n$\emph{. Then a solution of \eqref{qqq} exists if and only if the
volume of }$M$\emph{ is quantized to belong to the invariant }$q_{M}\subset
Z$\emph{ defined as the subset of }$Z$\emph{ }%
\begin{align}
&  q_{M}=\left\{  \mathrm{deg}(\phi_{+})+\mathrm{deg}(\phi_{-})\mid\phi_{\pm
}:M\rightarrow S^{n}\right\}  ,\nonumber\\
&  |\phi_{+}|(x)+|\phi_{-}|(x)\neq0,\,\forall x\in M,
\end{align}
\emph{where }$deg$\emph{ is the topological degree of the smooth maps and
}$|\phi|(x)$\emph{ is the Jacobian of }$\phi$\emph{ at }$x\in M$\emph{.}

The invariant $q_{M}$ makes sense in any dimension. For $n=2,3$, and any $M$,
it contains all sufficiently large integers. The case $n=4$ is much more
difficult, but the proof that it works for all spin manifold is given in
\cite{CCM1}.

It is natural from the point of view of differential geometry, to consider the
two sets of $\Gamma$-matrices and then take the operators $Y$ and $Y^{\prime}$
as being the correct variables for a first shot at a theory of quantum
gravity. Once we have the $Y$ and $Y^{\prime}$ we can use them and get a map
$(Y,Y^{\prime}):M\rightarrow S^{n}\times S^{n}$ from the manifold $M$ to the
product of two $n$-spheres. Given a compact $n$-dimensional manifold $M$ one
can find a map $(Y,Y^{\prime}):M\rightarrow S^{n}\times S^{n}$ which embeds
$M$ as a submanifold of $S^{n}\times S^{n}$. This is a known result, the
strong embedding theorem of Whitney, \cite{WW}, which asserts that any smooth
real $n$-dimensional manifold (required also to be Hausdorff and
second-countable) can be smoothly embedded in the real 2n-space. Of course
${\mathbb{R}}^{2n}={\mathbb{R}}^{n}\times{\mathbb{R}}^{n}\subset S^{n}\times
S^{n}$ so that one gets the required embedding. This result shows that there
is no restriction by viewing the pair $(Y,Y^{\prime})$ as the correct
\textquotedblleft coordinate" variables.

\emph{Quantization of four-volume.---}
We now specialize to a four-dimensional Euclidean manifold and for simplicity
consider only one set of maps, as this does not affect the analysis, and write
$Y=Y^{A}\Gamma_{A},\qquad A=1,2,...,5,$ where $Y^{A}$ are real and $\Gamma
_{A}$ are the Hermitian gamma matrices satisfying $\left\{  \Gamma^{A}%
,\Gamma^{B}\right\}  =2\delta^{AB}$. The condition $Y^{2}=1$ implies
\begin{equation}
Y^{A}Y^{A}=1, \label{4}%
\end{equation}
where the index $A$ is raised and lowered with $\delta^{AB},$ thus defining
the coordinates on the sphere $S^{4}.$ Notice that $Y^{A}$ are functions on
the Euclidian manifold $M_{4}$ which depend on the coordinates $x^{\mu}.$ The
Dirac operator on $M_{4}$ is
\begin{equation}
D=\gamma^{\mu}\left(  \frac{\partial}{\partial x^{\mu}}+\omega_{\mu}\right)  ,
\label{5}%
\end{equation}
where $\gamma^{\mu}=e_{a}^{\mu}\gamma^{a}$ and $\gamma^{1}\gamma^{2}\gamma
^{3}\gamma^{4}=\gamma,$ and $\omega_{\mu}$ is the connection, so that $\left[
D,Y\right]  =\gamma^{\mu}\frac{\partial Y^{A}}{\partial x^{\mu}}\Gamma_{A}.$
Using the properties of gamma matrices one can check that the condition
(\ref{yyy}) reduces to
\begin{equation}
\det\left(  e_{\mu}^{a}\right)  =\frac{1}{4!}\epsilon^{\mu\nu\kappa\lambda
}\epsilon_{ABCDE}Y^{A}\partial_{\mu}Y^{B}\partial_{\nu}Y^{C}\partial_{\kappa
}Y^{D}\partial_{\lambda}Y^{E}. \label{6}%
\end{equation}
Integrating over the volume of the manifold we find that%
\begin{equation}%
\begin{array}
[c]{c}%
V=\int\frac{1}{4!}\epsilon^{\mu\nu\kappa\lambda}\epsilon_{ABCDE}Y^{A}%
\partial_{\mu}Y^{B}\partial_{\nu}Y^{C}\partial_{\kappa}Y^{D}\partial_{\lambda
}Y^{E}d^{4}x\\
=%
{\displaystyle\int}
\det\left(  e_{\mu}^{a}\right)  d^{4}x=\frac{8\pi^{2}w}{3}.
\end{array}
\label{7}%
\end{equation}
This number $w$ is the number of components when using the one-sided equation
\eqref{yyy} but using \eqref{jjj} one gets the sum of the degrees of the maps
$Y_{\pm}:M\rightarrow S^{n}$ \cite{Greub}. Thus, we conclude that in
noncommutative geometry the volume of the compact manifold is quantized in
terms of Planck units. This solves a basic difficulty of the spectral action
\cite{AC} whose huge cosmological term is now quantized and no longer
contributes to the field equations.

\emph{Gravitational action and cosmological constant.---}
Let us study consequences of the four volume quantization for Einstein
gravity. For simplicity we shall utilize one set of maps $Y^{A}\left(
x\right)  $ since most of the details of what follows do not change when two
sets $Y_{\pm}^{A}\left(  x\right)  $ are used instead. First we consider
Euclidian compact spacetime and implement the kinematic constraints (\ref{4})
and (\ref{6}) in the action for gravity through Lagrange multipliers. This
action then becomes
\begin{align}
I  &  =-\frac{1}{2}%
{\displaystyle\int}
d^{4}x\sqrt{g}R+\frac{1}{2}%
{\displaystyle\int}
d^{4}x\sqrt{g}\lambda^{\prime}\left(  Y^{A}Y^{A}-1\right)  +\nonumber\\
&
{\displaystyle\int}
d^{4}x\frac{\lambda}{2}\left(  \sqrt{g}-\frac{\epsilon^{\mu\nu\kappa\lambda}%
}{4!}\epsilon_{ABCDE}Y^{A}\partial_{\mu}Y^{B}\partial_{\nu}Y^{C}%
\partial_{\kappa}Y^{D}\partial_{\lambda}Y^{E}\right)  , \label{8a}%
\end{align}
where $8\pi G=1.$ Notice that the last term is a four-form and represents the
volume element of a unit four-sphere. It can be written in differential forms
and is independent of variation of the metric. Variation of the action with
respect to the metric gives%
\begin{equation}
G_{\mu\nu}+\frac{1}{2}g_{\mu\nu}\lambda=0, \label{10}%
\end{equation}
where $G_{\mu\nu}=R_{\mu\nu}-\frac{1}{2}Rg_{\mu\nu}$ is the Einstein tensor.
Tracing this equation gives $\lambda=-\frac{1}{2}G,$ and as a result equations
for the gravitational field become traceless
\begin{equation}
G_{\mu\nu}-\frac{1}{4}g_{\mu\nu}G=0. \label{12}%
\end{equation}
Using the Bianchi identity these equations imply that $\partial_{\mu}G=0,$ and
hence $G=4\Lambda,$ where $\Lambda$ is the cosmological constant arising as a
\textit{constant of integration }(compare to \cite{HT}). Variation of the
action with respect to $Y^{A}$ does not lead to any new equations because the
last term in equation (\ref{8a}) is a topological invariant if $Y^{A}Y^{A}=1.$

One immediate application is that, in the path integration formulation of
gravity, and in light of having only the traceless Einstein equation
(\ref{12}), integration over the scale factor is now replaced by a sum of the
winding numbers with an appropriate weight factor. We note that for the
present universe, the winding number equal to the number of Planck quanta
would be $\sim10^{61}$ \cite{CC}

\emph{Three-volume quantization and mimetic matter.---}
In reality spacetime is Lorentzian and generically has one noncompact
dimension corresponding to time. Therefore, the condition for the volume
quantization is literally non-applicable there. However being implemented in
the Euclidian action it leads nevertheless to the appearance of the
cosmological constant as an integration constant even in the Lorentzian
spacetime. To show this let us first make a Wick rotation and then
decompactify $M_{4}$ to $\mathbb{R\times}S^{3}.$ With this purpose we set
$Y^{5}=\eta X$ and one of the coordinates say $x^{4}\rightarrow\eta t$ and
take the limit $\eta\rightarrow0$. In this limit equation (\ref{4}) becomes
$Y^{a}Y^{a}=1,\qquad a=1,\cdots,4$, and the constraint (\ref{6}) turns to
\begin{align}
\sqrt{g}  &  =\lim_{\eta\rightarrow0}\left(  \frac{1}{4!}\kappa^{4}%
\epsilon^{\mu\nu\kappa\lambda}\epsilon_{ABCDE}Y^{A}\partial_{\mu}Y^{B}%
\partial_{\nu}Y^{C}\partial_{\kappa}Y^{D}\partial_{\lambda}Y^{E}\right)
\nonumber\\
&  =\frac{1}{3!}\epsilon^{\mu\nu\kappa\lambda}\epsilon_{abcd}\left(
\partial_{\mu}X\right)  Y^{a}\partial_{\nu}Y^{b}\partial_{\kappa}Y^{c}%
\partial_{\lambda}Y^{d}. \label{16}%
\end{align}

The Lorentzian action for the gravity is%
\begin{equation}
I=-\frac{1}{2}%
{\displaystyle\int}
d^{4}x\sqrt{-g}R+\frac{1}{2}%
{\displaystyle\int}
d^{4}x\sqrt{-g}\lambda^{\prime}\left(  Y^{a}Y^{a}-1\right) \nonumber
\end{equation}%
\begin{equation}
+%
{\displaystyle\int}
d^{4}x\frac{\lambda}{2}\left(  \sqrt{-g}-\frac{1}{3!}\epsilon^{\mu\nu
\kappa\lambda}\epsilon_{abcd}\left(  \partial_{\mu}X\right)  Y^{a}%
\partial_{\nu}Y^{b}\partial_{\kappa}Y^{c}\partial_{\lambda}Y^{d}\right)  .
\label{3d}%
\end{equation}
The equations of motion in this case are the same as before and the
cosmological constant arises as a constant of integration. The variable $X$ in
(\ref{3d}) is \emph{a priori} unrestricted. We will show now that the
requirement of the volume quantization of $S^{3}$ in the mapping
$M_{4}\rightarrow\mathbb{R\times}S^{3}$ leads to the following normalization
condition for this variable,
\begin{equation}
g^{\mu\nu}\partial_{\mu}X\partial_{\nu}X=1. \label{19}%
\end{equation}
Let us consider the $3+1$ splitting of space-time, so that
\begin{equation}
ds^{2}=h_{ij}\left(  dx^{i}+N^{i}dt\right)  \left(  dx^{j}+N^{j}dt\right)
-N^{2}dt^{2} \label{20}%
\end{equation}
where $N\left(  x^{i},t\right)  $ and $N^{i}\left(  x^{i},t\right)  $ are the
lapse and shift functions respectively and $\sqrt{-g}=N\sqrt{h}.$ Consider a
hypersurfaces $\Sigma$ defined by constant $t.$ Taking
\begin{equation}
\partial_{i}X=0,\qquad\partial_{t}X=N, \label{22}%
\end{equation}
which satisfy (\ref{19}) we have (for details see \cite{CCM2})
\begin{equation}
\left(  N\sqrt{h}\right)  _{\Sigma}=\frac{1}{3!}N\left(  \epsilon^{ijk}%
Y^{a}\partial_{i}Y^{b}\partial_{j}Y^{c}\partial_{k}Y^{d}\epsilon
_{abcd}\right)  , \label{23}%
\end{equation}
and therefore $%
{\displaystyle\int\limits_{\Sigma}}
\sqrt{h}d^{3}x=w(\frac{4}{3}\pi^{2}),$ where $\ w$ is the winding number for
the mapping $\Sigma\rightarrow S^{3}.$ Thus we have shown that (\ref{19})
implies quantization of the volume of of compact 3d hypersurfaces in 4d
spacetime. This condition can be understood as a restriction of the maps
$Y^{A}\left(  x\right)  $ along directions orthogonal to the hypersurface
$\Sigma$, to be length-preserving. To incorporate this condition in the action
(\ref{3d}) we add to it the term%
\begin{equation}
+\int d^{4}x\sqrt{-g}\lambda^{\prime\prime}\left(  g^{\mu\nu}\partial_{\mu
}X\partial_{\nu}X-1\right)  \label{24a}%
\end{equation}
which corresponds to mimetic dark matter \cite{CM,CMV}. Thus the resulting
action describes both dark matter and dark energy. Both substances arise
automatically when the kinematic 4d and 3d compact volume-quantization in
noncommutative geometry is incorporated in the gravity action. We note that
the same field $X$ could be used when we consider two different set of maps
$Y_{+}^{A}\left(  x\right)  $ and $Y_{\_}^{A}\left(  x\right)  $ from $\Sigma$
to $S^{3}.$

\emph{Area quantization and black holes.---}
We can determine the conditions under which the area of any compact 2d
submanifold of the 4d manifold must also be quantized. One can show (see
\cite{CCM2} for details) that by writing $Y^{4}=\eta X^{1},$ $Y^{5}=\eta
X^{2}$ and rescaling the coordinates transverse to the 2d hypersurface
$\Sigma$ as $x^{\alpha}\rightarrow\eta x^{\alpha},$ we get%
\begin{equation}
\sqrt{^{\left(  2\right)  }g}=\det\left(  e_{i}^{a}\right)  =\frac{1}%
{2!}\epsilon^{ij}\epsilon_{ABC}Y^{A}\partial_{i}Y^{B}\partial_{j}Y^{C},
\end{equation}
provided that the area preserving condition on the hypersurface
\begin{equation}
\det\left(  g^{\mu\nu}\partial_{\mu}X^{m}\partial_{\nu}X^{n}\right)
|_{\Sigma}=1,\qquad m,n=1,2,
\end{equation}
is satisfied, where the index $A$ in $Y^{A}$ now $=1,2,3$ and $x^{i}$ are
coordinates on the hypersurface . Hence, the area of a two-dimensional
manifold is quantized%
\begin{align}
S  &  =%
{\displaystyle\int}
\sqrt{^{\left(  2\right)  }g}d^{2}x\nonumber\\
&  =\int\frac{1}{2!}\epsilon^{\mu\nu}\epsilon_{ABC}Y^{A}\partial_{\mu}%
Y^{B}\partial_{\nu}Y^{C}d^{2}x=4\pi n,
\end{align}
where $n$ is the winding number for the mapping of two-dimensional manifold to
the sphere $S^{2},$ defined by $Y^{A}Y^{A}=1.$

This can have far-reaching consequences for black holes and de Sitter space.
In particular, the area of the black hole horizon must be quantized in units
of the Planck area (see also \cite{BM1}). Because the areas of the black hole
of mass $M$ is equal to $A=16\pi M^{2},$ this implies the mass quantization
$M_{n}=\frac{\sqrt{n}}{2}.$

As it was shown in \cite{BM2} the Hawking radiation in this case can be
considered as a result of quantum transitions from the level $n$ to the nearby
levels $n-1,n-2,...$ As a result even for large black holes the Hawking
radiation is emitted in discrete lines and the spectrum with the thermal
envelope is not continuous. The distance between the nearby lines for the
large black holes is of order
\begin{equation}
\omega=M_{n}-M_{n-1}\simeq\frac{1}{4\sqrt{n}}=\frac{1}{8M}, \label{27}%
\end{equation}
and proportional to the Hawking temperature, while the width of the line is
expected to be at least ten times less than the distance between the lines
\cite{BM2}. Note that taking the minimal area to be $\alpha$ larger than the
Planck area changes the distance between the lines by a factor $\alpha.$
Within the framework discussed, area quantization could thus have observable
implications for evaporating black holes. Applying the same reasoning to the
event horizon in de Sitter universe we find that the cosmological constant in
this case must be quantized as $\Lambda_{n}=\frac{3}{n}.$ It is likely that
this quantization can have drastic consequences for the inflationary universe,
in particular, in the regime of self-reproduction (see also \cite{Gomez}).

\begin{acknowledgments}
AHC is supported in part by the National Science Foundation under Grant No.
Phys-1202671. The work of VM is supported by TRR 33 \textquotedblleft The Dark
Universe\textquotedblright\ and the Cluster of Excellence EXC 153
\textquotedblleft Origin and Structure of the Universe\textquotedblright
\end{acknowledgments}

\end{document}